\documentclass[
    ,final            % use final for the camera ready runs
]
  {aipproc}

\layoutstyle{6x9}

\newcommand{\tr}{{\textrm {tr}}}
\newcommand{\htr}{{\widehat\tr}}
\newcommand{\Tr}{{\textrm {Tr}}}
\newcommand{\g}{{\textrm {g}}}
\newcommand{\bnu}{{\overline \nu}}
\newcommand{\hnu}{{\widehat \nu}}
\newcommand{\bfx}{{\mathbf x}}
\newcommand{\D}{{\widehat D}}
\newcommand{\E}{{\widehat E}}
\newcommand{\F}{{\widehat F}}
\newcommand{\cD}{{\mathcal D}}
\newcommand{\cL}{{\mathcal L}}
\newcommand{\Om}{{\widehat \Omega}}
\newcommand{\MS}{\overline{\textrm{MS}}}
\newcommand{\ha}{a^{T}}
\newcommand{\thruu}[1]{\mathrel{\mathop{#1\!\!\!\!\!/}}}
\newcommand{\half}{{\textstyle\frac{1}{2}}}

\begin{document}

\title{One-loop effective action of QCD at high temperature using the heat kernel method\thanks{Presented at the IX Hadron Physics and VII Relativistic Aspects of Nuclear Physics, Angra dos Reis, Rio de Janeiro, Brazil, March 28 to April 03, 2004.}}

\author{E. Meg\'{\i}as}{
address={Departamento de F{\'{\i}}sica Moderna,  
Universidad de Granada. 18071-Granada (Spain)}
}

\begin{abstract}
Perturbation theory is an important tool to describe the properties of QCD at very high temperatures. Recently a new technique has been proposed to compute the one-loop effective action of QCD at finite temperature by making a gauge covariant derivative expansion, which is fully consistent with topologically small and large gauge transformations (also time dependent transformations) \cite{megias2}. This technique is based on the heat kernel expansion, and the thermal Wilson line plays an essential role \cite{megias1}. We consider a general SU($N_c$) gauge group.
\end{abstract}

\maketitle

%%%%%%%%%%%%%%%%%%%%%%%%%%%%%%%%%%%%%%%%%%%%
%% MAINMATTER
%%%%%%%%%%%%%%%%%%%%%%%%%%%%%%%%%%%%%%%%%%%%

\vspace{-0.25cm}

%%%%%%%%%%%%%%%%%%%%%%%%%%%%%%
%%%%%%%%%%%%%%%%%%%%%%%%%%%%%%
{\bf Introduction.} The effective action (EA) plays a prominent role in quantum field theo\-ry, since it embodies the renormalized properties of the system. To one loop it takes the form $c\Tr\log(K)$, where $K$ is the differential operator controlling the quadratic quantum fluctuations above a classical background. This quantity is afflicted by ultraviolet divergences, and it is useful to express it by means of a proper time representation \cite{schwinger},
\begin{equation}
-\Tr\log(K) = \int_0^\infty \frac{d\tau}{\tau}\, \Tr \, e^{-\tau K} 
= \int_0^\infty \frac{d\tau}{\tau} \int d^{D}x \,\tr \,\langle x|e^{-\tau K}|x\rangle \,.
\end{equation}
``$\tr$'' refers to trace in internal spaces. The (diagonal) heat kernel $\langle x|e^{-\tau K}|x\rangle$ is UV finite.

The heat kernel at finite temperature has been computed for a Klein-Gordon operator,\footnote{We consider Klein-Gordon operators of the form, $K= M(x)-D_\mu^2 $, where $D_\mu= \partial_\mu-i\g A_\mu(x)$ is the covariant derivative, $M(x)$ is a scalar (external) field, and $A_\mu(x)$ are the gauge fields.} through the so called heat kernel expansion, in the completely general case of non Abelian and non stationary gauge fields and external fields \cite{megias2,megias1}. It is of the form 
\begin{equation}
\langle x|e^{-\tau(M-D_\mu^2)}| x\rangle= 
(4\pi\tau)^{-D/2}\sum_n
\ha_n(x)\tau^n \,.
\label{eq:18b}
\end{equation}
It is an expansion in local and gauge covariant operators. The so called Seeley-DeWitt coefficients, $\ha_n$, are contructed with operators of mass dimesion $2n$. The untraced Polyakov loop or thermal Wilson line plays a fundamental role in maintaining manifest gauge invariance at each order. It appears inside $\ha_n$, and it is defined as 
\begin{equation}
\Omega(x)= T\exp\left(i\g\int_{x_0}^{x_0+\beta}
A_0(x_0^\prime,\bfx)dx_0^\prime\right) \,.
\end{equation}

%%%%%%%%%%%%%%%%%%%%%%%%%%%%%%
%%%%%%%%%%%%%%%%%%%%%%%%%%%%%%
{\bf Effective action of QCD.} In Ref.~\cite{megias2} the heat kernel has been applied to obtain the 1-loop EA of QCD at high temperature. The Euclidean partition function of QCD is
\begin{equation}
Z[A,\overline{q},q]= \int \cD A 
\prod_{\alpha=1}^{N_f}
\cD\overline{q}_\alpha \cD q_\alpha
\exp\left\{ -\frac{1}{2}\int d^4x\,\tr(F_{\mu\nu}^2) 
- \int d^4x \,\sum_{\alpha=1}^{N_f}\overline{q}_\alpha \,i\thruu{D} \,q_\alpha 
\right\}\,,
\label{eq:42}
\end{equation}
where $ F_{\mu\nu}=i\g^{-1}[D_\mu,D_\nu]$ is the field strength. The gauge group of color is SU($N_c$).

The full EA can be obtained upon functional integration of the quark and gluon fields. The quark contribution to the EA at 1-loop is [convention $Z=e^{-\Gamma[A]}$]  
\begin{equation}
\Gamma_q[A]= -\frac{N_f}{2}\Tr\,\log(-\thruu{D}{}^2) =: 
\int d^4x \cL_q(x) \,. 
\label{eq:q1}
\end{equation}
For the gluon fields we use the background field method, in which the gauge field is split into a classical background field plus a quantum fluctuation, i.e. $A_\mu \rightarrow A_\mu + a_\mu$, in~\eqref{eq:42}. The standard procedure consists of adding a gauge fixing term and the corresponding Fadded-Popov term in the action. The gluon contribution is then 
\begin{equation}
\Gamma_g[A]= \frac{1}{2}\Tr\log\left(-\delta_{\mu\nu}\D_\lambda^2+2i\g\F_{\mu\nu} \right)
-\Tr\log\left(-\D_\mu^2\right)
=: \int d^4x \cL_g(x)\,.
\label{eq:g1}
\end{equation}
The operators in \eqref{eq:q1} and \eqref{eq:g1} act in the fundamental and adjoint representations, respectively. They are of the Klein-Gordon form, so we can use the proper time repre\-sentation.\footnote{The integrals over $\tau$ are one-valued functions of the Polyakov loop, so gauge invariance is manifest.} For the different orders of the heat kernel expansion, we obtain 
\begin{eqnarray}
\cL_{0}(x) &=& \frac{\pi^2 T^4}{3} \left(
\frac{1+2N_cN_f-N_c^2}{15}
-\frac{N_f}{4}\tr\left[(1-4\overline\nu^2)^2\right] 
+2\htr\left[ \hnu^2(1-\hnu)^2 \right]  \right)  \,,
\\
\cL_2(x) &=&
\left(\frac{1}{2} -\g^2\beta_0 \log(\mu/4\pi T)
-\frac{\g^2 N_c}{6(4\pi)^2} \right) 
\,\tr(F_{\mu\nu}^2) 
\label{eq:3.31}
\\
&& +\frac{11\g^2}{12(4\pi)^2}\htr \left[
 \big( \psi(\hnu)+\psi(1-\hnu) \big)
\F_{\mu\nu}^2
\right]
\nonumber \\
&&-\frac{\g^2N_f}{3(4\pi)^2}\tr \left[
\left(\psi(\half+\bnu)+\psi(\half-\bnu) \right)F_{\mu\nu}^2
\right]
+\frac{2\g^2}{3(4\pi)^2}(N_c-N_f)  
\tr \left[ E_i^2 \right]
\,,
\nonumber \\
\cL_{3}(x) &=& -\frac{2\g^2}{(4\pi)^4 } \frac{N_f}{T^2}
 \tr \bigg[
\Big(\psi^{\prime\prime}(\half+\overline\nu) +\psi^{\prime\prime}(\half-\overline\nu) \Big)
\Big( 
\frac{1}{60}[D_\mu, F_{\mu\nu}]^2
-\frac{1}{24}[D_\lambda, F_{\mu\nu}]^2
\nonumber \\
&&+\frac{8}{45}i\g F_{\mu\nu}F_{\nu\lambda}F_{\lambda\mu}
-\frac{1}{20}[D_0,F_{\mu\nu}]^2
+\frac{1}{30}[D_i,E_i]^2
+\frac{1}{15}i\g E_iF_{ij}E_j
\Big)
\bigg]
\nonumber \\
&& +\frac{\g^2}{2(4\pi)^4} \frac{1}{T^2}
\htr \bigg[
\Big(\psi^{\prime\prime}(\hnu) 
+ \psi^{\prime\prime}(1-\hnu) \Big)
\Big( 
\frac{1}{30}[\D_\mu,\F_{\mu\nu}]^2
-\frac{1}{3}[\D_\lambda,\F_{\mu\nu}]^2
\nonumber \\
&&+\frac{61}{45}i\g\F_{\mu\nu}\F_{\nu\lambda}\F_{\lambda\mu}
-\frac{3}{5}[\D_0,\F_{\mu\nu}]^2
+\frac{1}{15}[\D_i,\E_i]^2
+\frac{2}{15}i\g\E_i\F_{ij}\E_j
\Big)
\bigg]
\,,
\end{eqnarray}
where $-\frac{1}{2} < \bnu < \frac{1}{2}$ and $0 < \hnu < 1$. \footnote{$\Omega(x)=e^{2\pi i\bnu}$ is in the fundamental representation, and $\Om(x)=e^{2\pi i\hnu}$ is in the adjoint one.} $\psi(q)$ is the digamma function. $E_i=F_{0i}$ is the electric field. $\g$ is the running coupling defined in the $\MS$ scheme. $\beta_0 = (11N_c-2N_f)/(3(4\pi)^2)$.

A new technique has been proposed recently for 1-loop QCD at high temperature \cite{diakonov}. It goes beyond ours in that all orders in $A_0$ are retained. But the authors use stationary configurations. Our results for $\cL_2$ agree for some terms that can be directly compared.

%%%%%%%%%%%%%%%%%%%%%%%%%%%%%%
%%%%%%%%%%%%%%%%%%%%%%%%%%%%%%
{\bf Dimensional reduced effective theory.} In the high temperature limit non stationary fluctuations become heavy and are therefore suppressed, and QCD behaves as an effective three-dimensional theory for the stationary configurations only. This effective theo\-ry is obtained by i) using stationary backgrounds and ii) taking purely non-stationary fluctuations only, that is, removing the static Matsubara mode. Doing this, one obtains\footnote{In that formulas we have rescaled $A_i$ and $A_0$ with different renormalization factors, so that $\cL_4^\prime$ looks like the zero temperature renormalized tree level:  $\; \g \rightarrow Z_g^{-1/2} \g \,, \;
 A_i \rightarrow Z_M^{1/2} A_i \,, \; A_0 \rightarrow Z_E^{1/2} A_0 $; 
with
\vspace{-0.3cm}
\begin{displaymath}
Z_M = Z_g = 1 +2\g^2\beta_0[\log(\mu/4\pi T)+\gamma_E]
-\frac{\g^2}{3(4\pi)^2} \left[-N_c+8 N_f\log 2\right]
\,, \quad
Z_E = Z_M -\frac{2\g^2}{3(4\pi)^2}(N_c-N_f)    
\,. 
\end{displaymath}
}
\begin{eqnarray}
\cL^\prime_0(\bfx) &=&
\g^2\left(\frac{N_c}{3}+\frac{N_f}{6}\right) T \langle A_0^2 \rangle
+\frac{\g^4}{4\pi^2 T} \langle A_0^2 \rangle^2
+\frac{\g^4}{12\pi^2 T}(N_c-N_f)\langle A_0^4 \rangle \,,
\nonumber \\
\cL^\prime_4(\bfx) &=& 
\frac{1}{2T} \langle F_{\mu\nu}^2 \rangle \,,
\\
\cL^\prime_6(\bfx) &=&
-\frac{2}{15}\frac{\g^2\zeta(3)}{(4\pi)^4 T^3}
\Bigg[
i\g(\frac{2}{3}N_c\!-\!\frac{14}{3}N_f)\langle F_{\mu\nu}F_{\nu\lambda}F_{\lambda\mu}\rangle
-(19 N_c\!-\!28 N_f)\langle [D_\mu,F_{\mu\nu}]^2\rangle
\nonumber \\
&&-(18 N_c \!- \!21 N_f) \langle [D_0,F_{\mu\nu}]^2\rangle
+(2 N_c \!- \!14 N_f) \langle [D_i,E_i]^2\rangle
+ 110\g^2 \langle A_0^2\rangle \langle F_{\mu\nu}^2 \rangle
\nonumber \\
&&
+i\g(4 N_c\!-\!28 N_f) \langle E_i F_{ij} E_j\rangle
\!+\!\g^2(110 N_c \!-\! 140 N_f) \langle A_0^2 F_{\mu\nu}^2\rangle
\!+\! 220 \g^2\langle A_0 F_{\mu\nu}\rangle ^2
\Bigg] \nonumber
\end{eqnarray}
where $\langle X \rangle := \tr(X)$, and $B_i = \frac{1}{2}\epsilon_{ijk}F_{jk}$ is the magnetic field. $\cL^\prime_6$ has been computed in~\cite{chapmanshort} for the gluon sector, and in \cite{wirstam} for the quark sector and for SU(3) in the absence of chromomagnetic field ($A_i = 0$). Our results agree with these.

%%%%%%%%%%%%%%%%%%%%%%%%%%%%%%%%%%%%%%%%%%%%
%% SAMPLE TABLE
%%
%% Shows the use of \tablehead and \tablenote
%% macros
%%%%%%%%%%%%%%%%%%%%%%%%%%%%%%%%%%%%%%%%%%%%

%%%%%%%%%%%%%%%%%%%%%%%%%%%%%%%%%%%%%%%%%%%%%%%%
%% BACKMATTER
%%%%%%%%%%%%%%%%%%%%%%%%%%%%%%%%%%%%%%%%%%%%%%%%

\vspace{0.4cm}
Work done in collaboration with E. Ruiz Arriola and L. L. Salcedo. Supported by funds provided by the Spanish DGI and FEDER Grant No. BFM2002-03218, Junta de Andaluc\'{\i}a Grant No. FQM-225, and EURIDICE Contract No. HPRN-CT-2002-00311.

%% If the bibliography is produced without BibTeX comment out the


%\bibliography{sample}

%\endinput

\end{document}